\documentstyle[11pt,newpasp,twoside,epsf]{article}
\def\edcomment#1{\iffalse\marginpar{\raggedright\sl#1\/}\else\relax\fi}
\reversemarginpar

\begin{document}
\title{Recycling of ghost galaxies: The origin of giant HI ring around NGC 1533}
\author{Kenji Bekki, Warrick J. Couch}
\affil{School of physics, University of New Southwales, Sydney, NSW, 2052, Australia
}
\author{Emma Ryan-Weber and  Rachel Webster}
\affil{School of Physics, University of Melbourne,VIC 3010, Australia}

\begin{abstract}
We propose that the giant HI ring recently discovered by HIPASS for S0 galaxy
NGC 1533 is formed by unequal-mass merging between gas-rich LSB (low surface brightness:
``ghost'') galaxies and HSB disks. The NGC 1533 progenitor HSB spiral is transformed
into a barred S0 during merging and the outer HI gas disk of the LSB is transformed 
into the giant HI ring. We also discuss two different possibilities  for the origin 
of isolated star-forming regions (``ELdot'' objects) in the giant gas ring.
\end{abstract}

\section{Ring formation from tidal disruption of LSBs}

Recent observational results of NGC 1533 HI rings and numerical models for this
objects are described in detail by Ryan-Weber et al. (2002) 
and by Bekki et al. (2003, in preparation),
respectively. We accordingly summarize briefly the important three results 
derived in this  numerical studies (See Figure 1 for the model details).
Firstly, the strong tidal filed of NGC 1533's progenitor HSB strips
the outer diffuse HI gas disk of the LSB and completely destroy the stellar disk
of the LSB. The stripped HI gas forms an incomplete gas ring and the LSB is swallowed by
the HSB finally.  
This outer HI ring formation is different from the merger scenario of 
polar ring formation
proposed by Bekki (1998). 
Secondly, the HSB forms a bar and loses its spiral arms owing to
tidal heating of the interaction, and therefore can be classified as a barred S0
after merging. Thirdly, star formation can proceed in the giant gas ring either
by the cloud-cloud collisions within the ring or by strong external gaseous
pressure of the hot gas surrounding the HI ring. 
The proposed merger models thus explain (1) the observed large total mass of the HI
ring, (2) structure and kinematics of the HI ring, and (3) the morphology of
the host S0 galaxy (NGC 1533). 
%The present model also predicts the presence of
%very diffuse stellar halo (around NGC 1533).
% which was formed from disruption
%of NGC 1533.

\begin{figure}
%\plotfiddle{sscf2.eps}{5.in}{0}{80.}{80.}{-300.}{500.}
\plotone{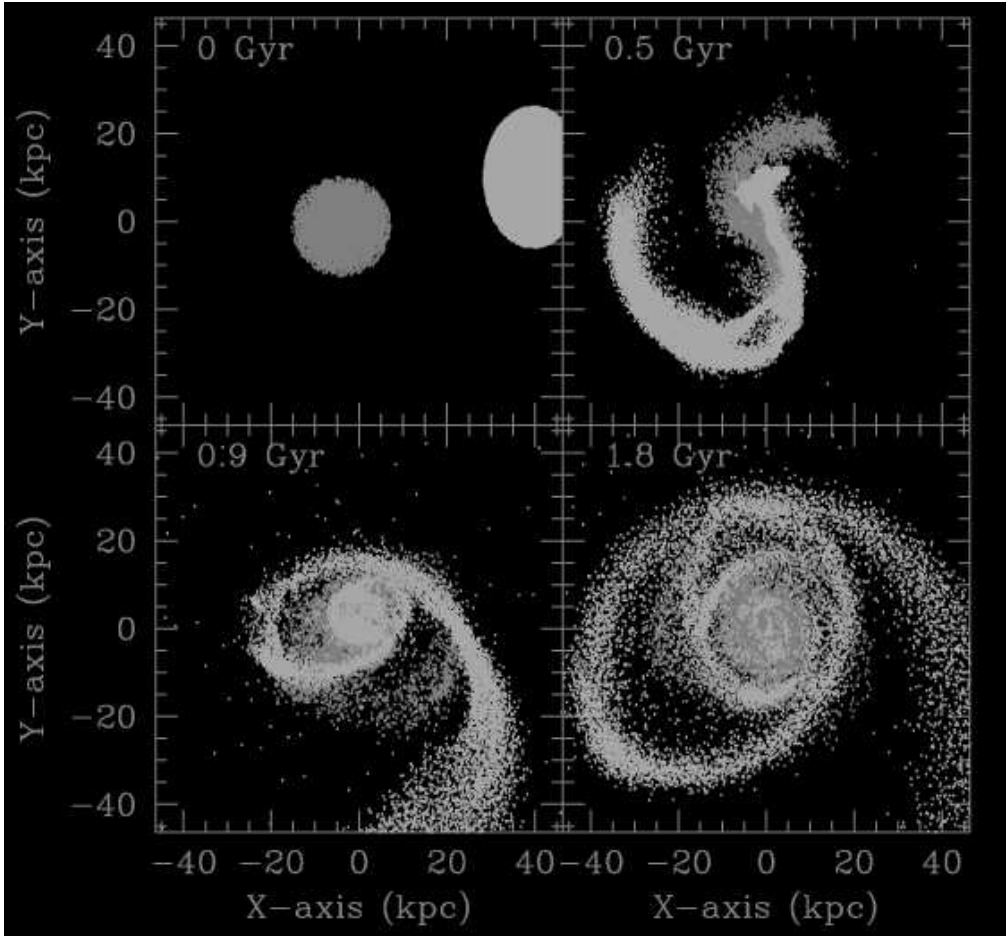}
%\plotone{sscf2.eps}
\caption{
Morphological evolution of the gas of the LSB (cyan) and
the stars of the HSB (magenta) projected onto the $x$-$y$ plane
(See astro-ph/ for color version of this figure).
Both disk are modeled based on the Fall-Efstathiou disk models,
and the LSB's central surface brightness is 1 mag fainter than
that of the HSB. The HI gas disk of the LSB is assumed to extend
2 times optical radius of the LSB. The total HI gas mass of
the LSB is assumed to be the same as that of the optical one.
The pericenter and the eccentricity of the orbit in this LSB-HSB
merger are set to be $0.5 \times R_{\rm d}$ and 0.8, respectively,
where $R_{\rm d}$ is the disk size of the HSB. TREESPH code
is used and an  isothermal equation of state is adopted
for the gas with a temperature of 
$7.3\times 10^3$ K (corresponding to a sound speed of 10 km $\rm s^{-1}$). 
Star formation (the Schmidt law) is also included in the SPH simulations.
As the two disks merge with each other, only the LSB disk (both gas and
stars) is ditroyed by the HSB's tidal field to form outer giant HI gas
rings. Since the rings originate from the outer gas disk of the LSB,
the rings are  composed mostly of gas and the total mass of the ring
can be estimated to be more than a few $10^9$ $\rm M_{\odot}$.
The shapes of the rings and the velocity field of
the HI gas ring reflect the initial LSB's orbit with respect
to the HSB. Some local regions along the ring have higher gas density
and thus  could be the possible formation sites of young stars,
which are observed as ''ELdots'' objects.
}
\end{figure}

\end{document}